# Secure Encrypted Virtualization is Unsecure!

**Zhao-Hui Du, Zhiwei(Victor) Ying, Zhenke Ma, Yufei(Michael) Mai,**
**Phoebe Wang, Jesse Liu and Jesse Fang***

**Tangram Technologies, Incorporated**
**Shanghai, China**
{zhaohui.du, victor.ying, zhenke.ma, michael.mai, phoebe.wang, jesse.liu, jesse.fang}@tangramtek.com

**Abstract**

Virtualization has become more important since cloud computing is getting more and more popular than before. There's an increasing demand for security among the cloud customers. AMD plans to provide Secure Encrypted Virtualization (SEV)[8] technology in its latest processor EPYC to protect virtual machines by encrypting its memory but without integrity protection. In this paper, we analyzed the weakness in the SEV design due to lack of integrity protection thus it is not so secure. Using different design flaw in physical address-based tweak algorithm to protect against ciphertext block move attacks, we found a realistic attack against SEV which could obtain the root privilege of an encrypted virtual machine protected by SEV. A demo to simulate the attack against a virtual machine protected by SEV is done in a Ryzen machine which supports Secure Memory Encryption (SME)[8] technology since SEV enabled machine is still not available in market.

## 1. Introduction

Virtualization[1] technology plays a more and more important role with the prevalent of cloud computing. Cloud provider provides hardware resources which allow multiple customers to share the hardware resources by launching different virtual machines (VM) in the shared platforms. Customers need neither to buy the hardware resources nor to maintain the systems by themselves which could greatly reduce their costs. VMs are isolated by a privileged software called hypervisor to provide secure environment for each customer. Intel and AMD have provided VT-X[11] and AMD-V (SVM)[12] technology respectively in their x86 CPU architecture to boost up virtualization performance with hardware support.

Hypervisor is responsible for managing multiple resources for VMs. For example, hypervisor allocates physical memories, creates virtual network device and sets up virtual IO port for a guest VM. It allows corresponding hardware resources efficiently shared and fully utilized. But there is still a secure gap in the virtualization technologies. Hypervisor could access all resources of VMs which grants cloud provider the ability to access private information of customers in the VMs. Some customers may be reluctant to use those virtualization technologies due to security concern.

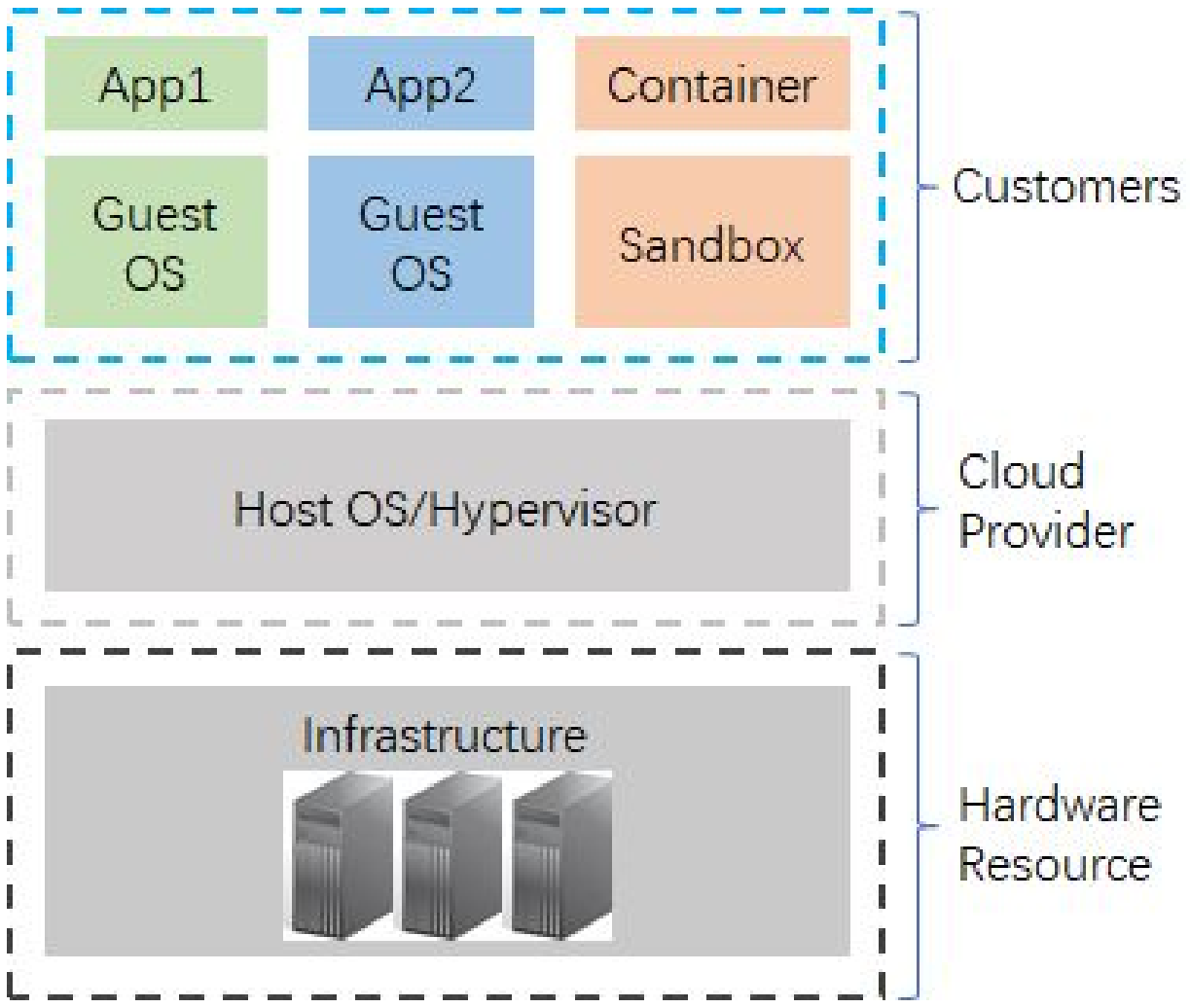

Picture 1. Virtualization

If there were bugs in the code of hypervisor, such as those have been found in reality[3~6], customers using the corresponding VMs are in the danger of losing their information.

To solve the problem, AMD provides SEV technology to enhance the SVM. Unlike Intel's Software Guard Extensions (SGX)[9] or Trust-zone[10] in ARM architecture which try to isolate trusted resources from untrusted access by special hardware logic, SEV encrypts memory of VMs by taking advantage of SME technology and encrypting different VMs with different keys to prevent unauthorized access. But SEV provides no integrity protection so that a malicious or compromised hypervisor is able to modify the cipher-text of VMs. This opens a door for potential attack. F. Hetzelt et al.[7] provided three proof-of-concept attacks against the SEV. Two of them used the unencrypted guest context and virtual machine control block (VMCB) and they were not available any more after the SEV Encrypted State (ES) feature was provided by AMD to solve the problem according to AMD latest Memory Encryption Whitepaper[14]. In the last proof-of-concept attack, a malicious hypervisor could launch a replay attack against a VM protected by SEV. But they agreed that SEV was still not broken in their paper.

AMD has used an additional physical address-based tweak algorithm to protect against cipher-text block move attacks. In this paper, we found out the tweak algorithm after analyzing it in a Ryzen machine which supports SME. The tweak algorithm is a linear function in the finite field $F_2$ if both input text and physical address are treated as vectors in $F_2$. Using the fact that there's no integrity protection of encrypted memory, we can design a real attack against SEV. A simulation of SEV environment was built in a Ryzen machine (which supports SME)since SEV is still not available in market. Attacker in hypervisor can obtain the root privilege of guest VM in tens of seconds.

## 2. Background

### 2.1 Secure Encrypted Virtualization

Secure Encrypted Virtualization (SEV)[13] was introduced by AMD to meet the security gap between existing virtualization technology and customer security requirements. SEV utilizes the Secure Memory Encryption (SME)[14] technology to encrypt memory contents of a guest VM. SME is a real time memory encryption technology. It makes the contents of the memory more resistant to memory snooping and cold boot attacks. The encryption key is manipulated by a "Security" Processor and is invisible to OS and application. SEV feature allows the memory contents of virtual machine(VM) to be transparently encrypted with key unique to each VM. The hypervisor could only read the cipher-text of the contents of VM. Similar to SME, all encryption keys are managed by "Security" Processor so that they're invisible to hypervisor.

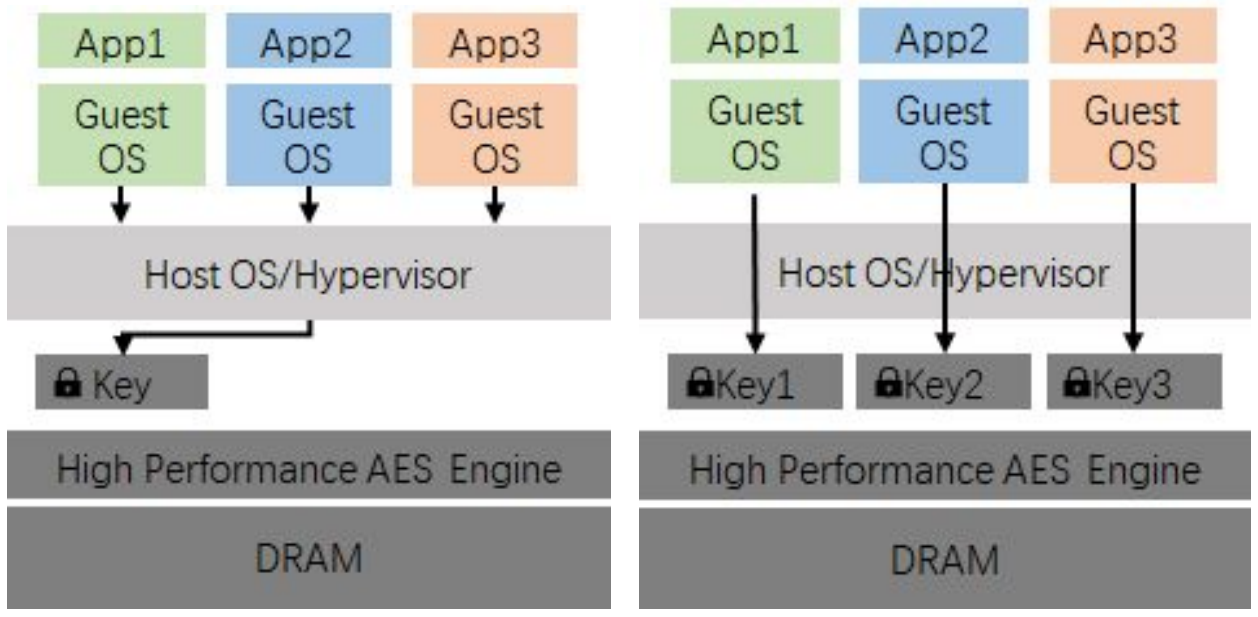

Picture 2 SME vs. SEV

The solution sounds perfect since all contents of VM have been encrypted by the well-known algorithm – AES[15] which provides confidentiality of VM. It looks like that although hypervisor can read and write to the cipher-text of VM, without the secret key, usually it doesn't know what the corresponding plaintext is and writing to the memory of VM is likely to make the VM crash and hypervisor can't get any secure information of the VM. But data integrity is also critical and D. Boneh[16] has emphasized the importance of integrity protection in his cryptography lecture.

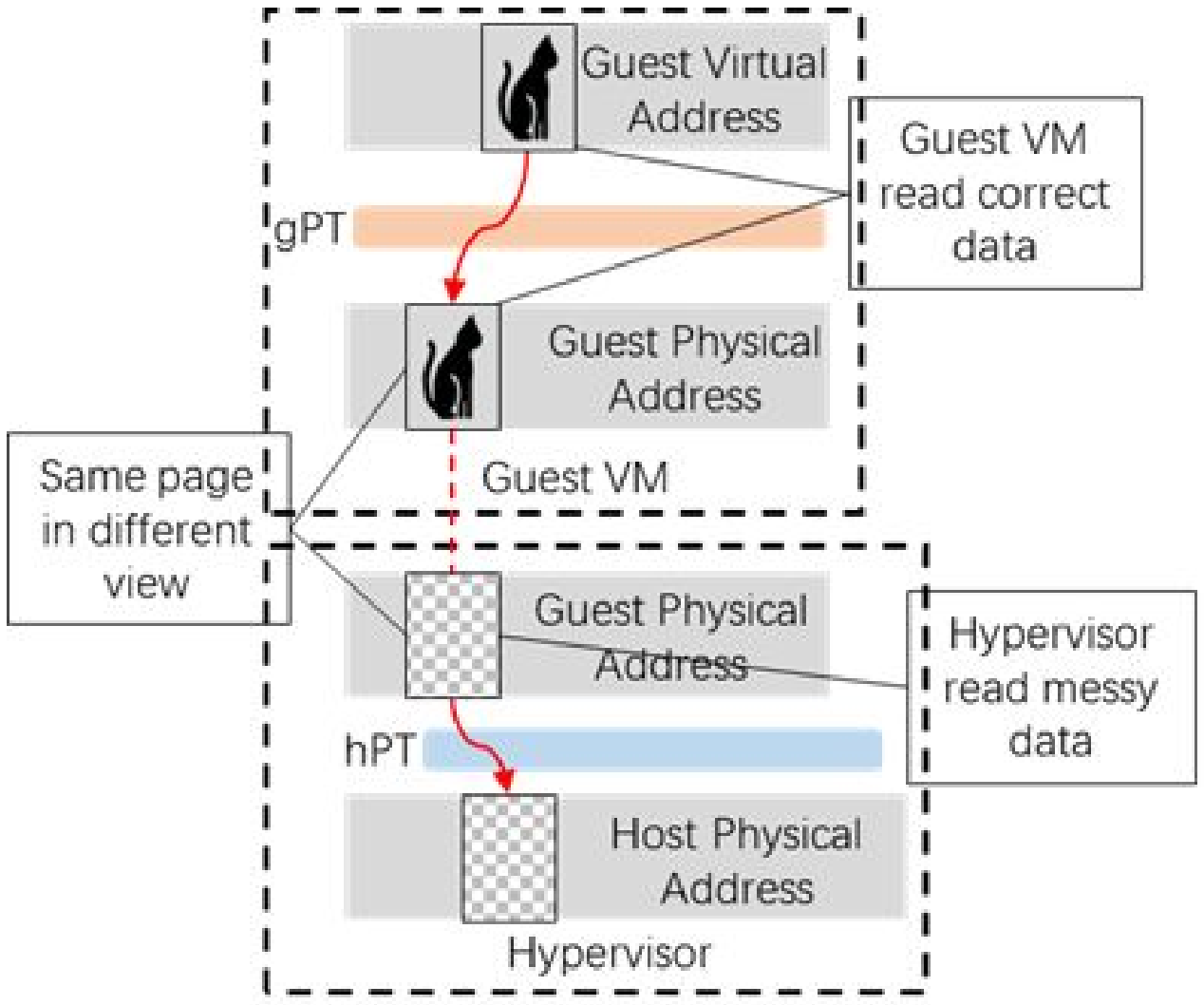

Picture 3 Hypervisor and guest VM

### 2.2 Paging in OS

In a computer system, each piece of memory has a unique address which is the physical address of the memory. To enable multiple applications to run in the same computer system simultaneously and share the limited memory resource, a memory management scheme is provided in most modern computer systems. The OS partitions the system memory into some blocks (usually with same size) called memory pages.

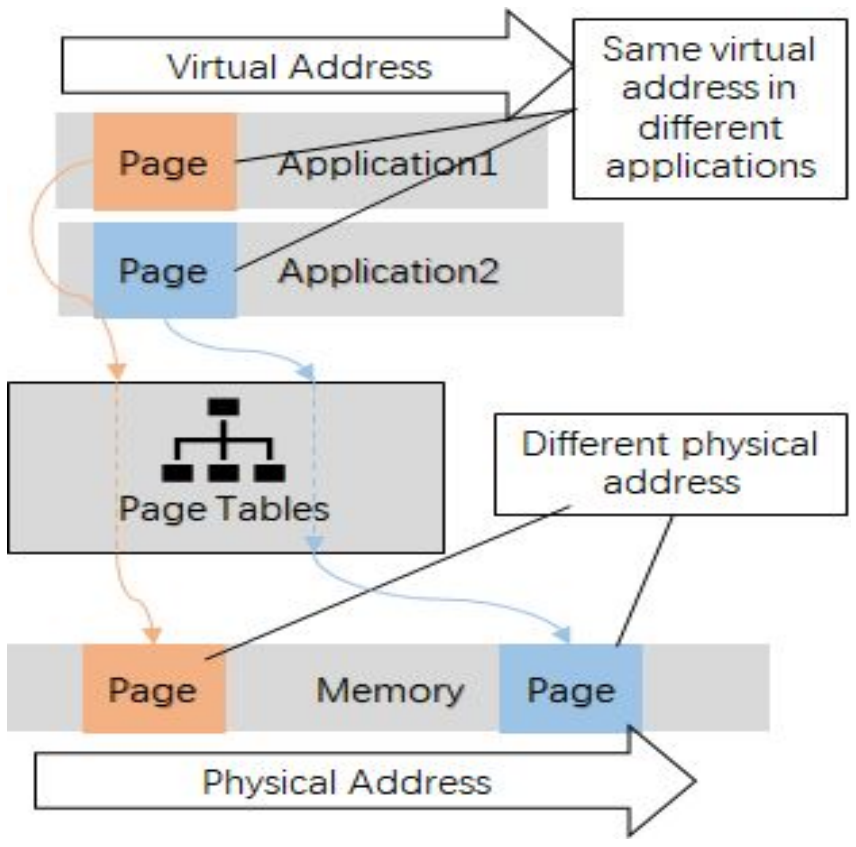

Picture 4 Virtual Address and Physical Address

OS could dynamically assign those memory pages to different applications and new application-specified addresses are assigned to those pages. The application-specified address is the virtual address of the memory page in the application. Page Tables are used to map the virtual address of each application to physical address or alternatively a location in external storage. A page table entry (PTE) is associated with each page and it also provides some extra attributes of the memory page.

## 2.3 Nested Page Table

To support virtualization, the hypervisor needs to assign physical memory pages dynamically to each VM. New addresses should be assigned to those pages inside the VM and VM will reassign those pages to applications running in the VM. The new address that hypervisor assigned to VM is the virtual machine physical address. OS in VM knows nothing about host physical address.

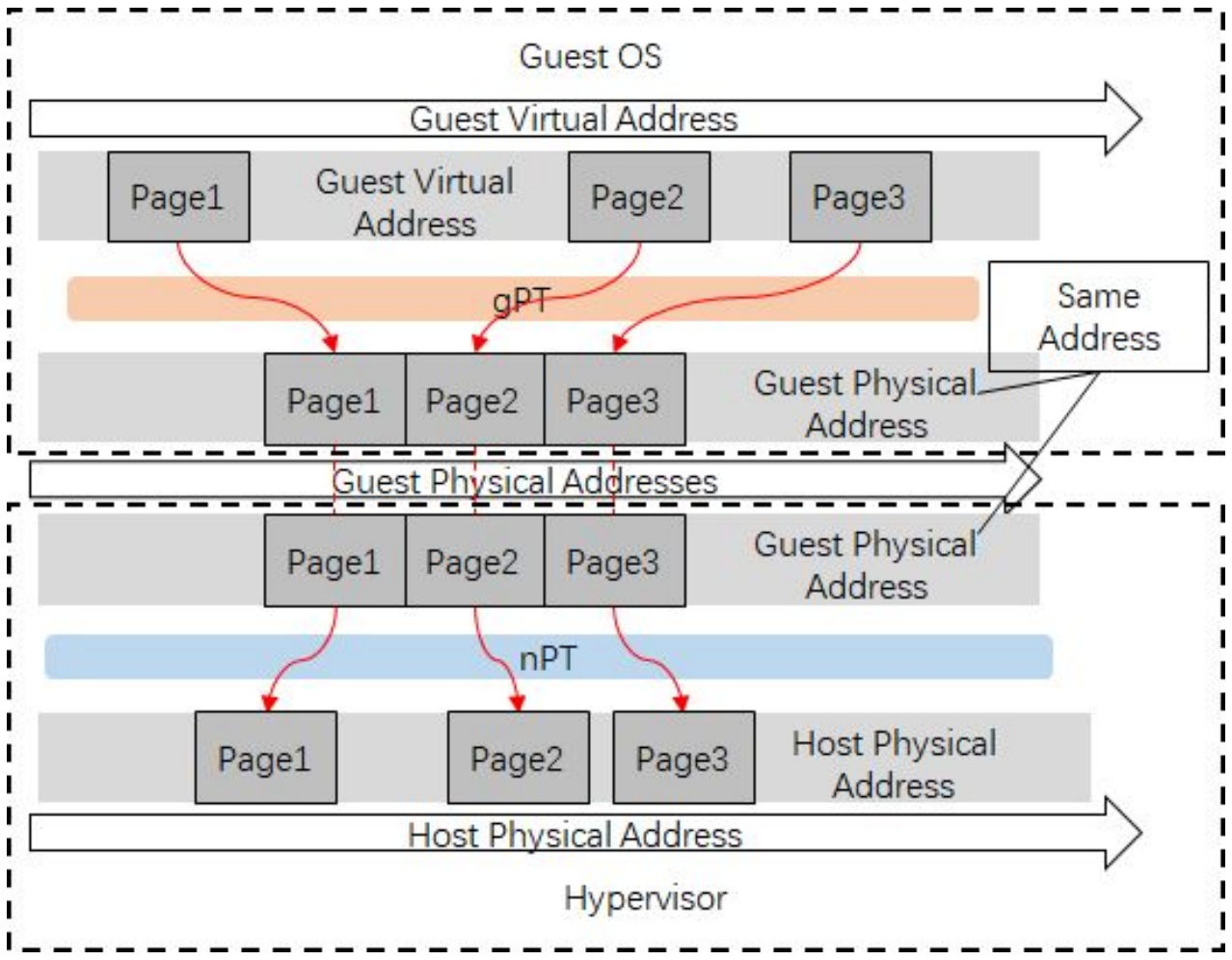

Picture 5 Nested Page Table in SVM

According to AMD64 Architecture Programmer Manual[18], Nested Page Table is provided. OS in virtual machine should maintain a guest page table (gPT) which maps virtual address to virtual machine physical address while hypervisor maintains a nested page table (nPT) which maps virtual machine physical address to host physical address.

## 2.4 The C-bit

SEV encrypts memory of different VMs by different keys and also hypervisor has its own key to encrypt its memory. But sometimes, VMs may need some non-encrypted memory. For instance, a VM may use shared memory to share data with hypervisor or a hardware. For both SME and SEV, A C-bit in page table entry of each page is used to indicate whether the page has been encrypted. When SEV is enabled, VM could use the C-bit in gPT to determine whether a page should be encrypted by the key unique to the VM. Page Table and codes are always treated as encrypted data even though the corresponding C-bit is not set.

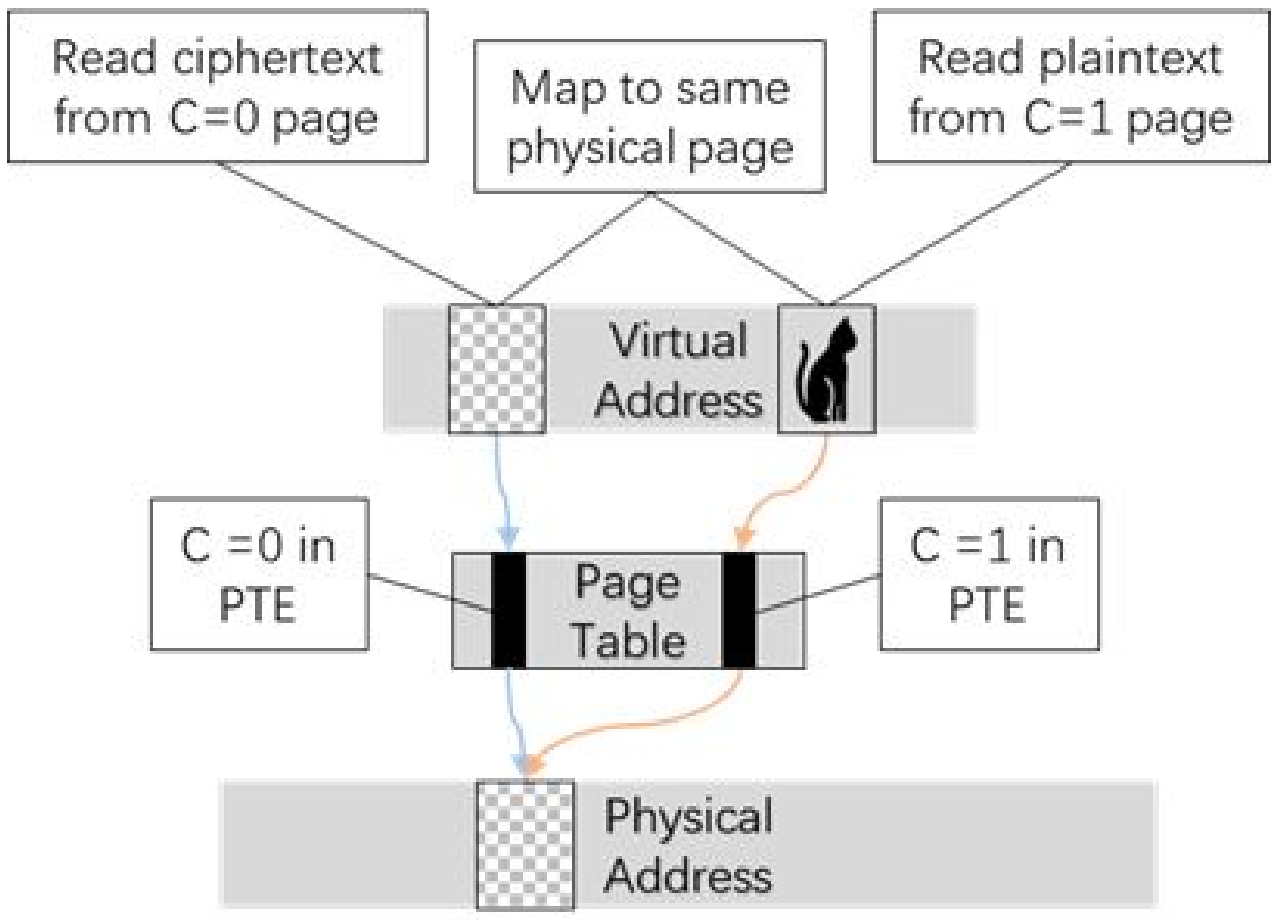

Picture 6 C-bit in PTE

In both SME and SEV, if two virtual addresses are mapped to same physical address but the C-bit is set for one virtual address and not set for another one, load access from first virtual address always gets the plaintext transparently and the load access from second virtual address always gets the cipher-text given the cache has been reflushed between the two loads. Similarly, store access to the first virtual address are treated as plaintext and encrypted transparently and store access to the second virtual address is not encrypted and the data is treated as cipher-text if it is loaded via the first virtual address later.

## 2.5 AES algorithm

SEV and SME use AES[15] as the memory encryption algorithm according to AMD's *Memory Encryption Whitepaper*. Encryption is a process to encode some user input -- the plaintext into cipher-text so that only authorized parties could decode the cipher-text into plaintext again. In symmetric-key encryption algorithm such as AES, all authorized parties share a secret key and both encryption and decryption could be processed only when the secret key is available. AES is a block cipher algorithm where 128-bits block is used with three different key lengths: 128, 196 and 256 bits.

There are multiple modes of operation available for block cipher. Modes like CBC, CFB, OFB chain different blocks together so that they are not suitable for memory encryption. Usually memory is accessed randomly and the performance penalty will be too heavy if those modes are used because multiple blocks of data will be involved to decode the data of a single block. CTR mode could provide better performance but the mode is not secure because each counter should be used only once unless the key has been changed in the viewpoint of security. Since no extra memory is used to save the counter, counter for each memory address must be reused and it violates the security requirement of CTR mode. So ECB mode is the only choice if taking into account both performance and security. There's a well-known secure issue that ECB mode generates same ciphertext output with same plaintext input. . Figure 7 depicts the problem that in the encrypted picture, we could still find the outline of human being head.

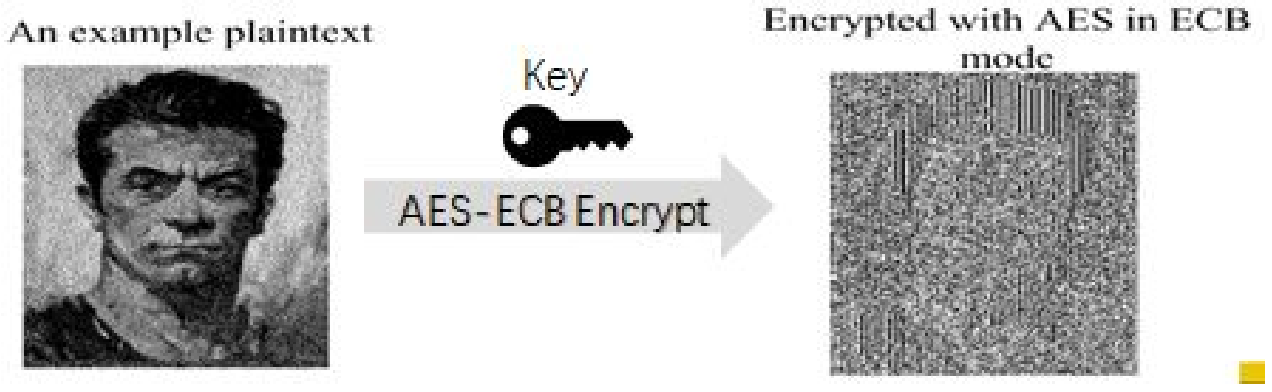

Figure 7. AES ECB mode secure issue

To overcome the problem, AMD uses a physical address-based tweak algorithm to combine physical address together with input plaintext. The output of the tweak algorithm is encrypted by AES algorithm again to generate the cipher-text. By applying the tweak algorithm, equal plaintexts in different physical addresses are encrypted into different cipher-texts.

## 3. Weakness of SEV

One of the problem in the SEV design is that the data integrity protection is not provided while it is critical to security design according to D. Boneh[16]. Missing of integrity protection means malicious hypervisor could modify the cipher-texts of a guest VM. A malicious hypervisor could also overwrite the data of any memory location with older version data from same location (replay attack). F. Hetzelt et al.[7] has introduced a proof-of-concept attack to launch a replay attack on VM.

Another problem is that after a VM is started, the VM Encryption Key (VEK) will never change. The life range of the encryption key will never end as long as the VM keeps alive. According to birthday paradox[19], when a key (the key length is 128-bits) is used for around $2^{64}$ times, the probability of existing equal cipher-texts becomes obvious. Equal cipher-texts indicates that there's a simple relationship between the corresponding plaintexts. If one of the plaintexts is known in advance, attacker could calculate another plaintext easily. To keep the probability of equal ciphertexts ignorable, it is suggested that an encryption key should be used for no more than $2^{48}$ times.

The third problem is that the nPT of nested page table is maintained by hypervisor and the hypervisor could modify the page table without permission from guest VM.

The last problem is that the physical address-based tweak algorithm uses the host physical address instead of virtual machine physical address. If a malicious attacker swaps the virtual machine physical address of two encrypted pages of the VM, VM is still able to decrypt both pages successfully but the data in the two pages has been exchanged without the permission from VM.

Attacks can be designed conceptually based on above problems. However it is not easy to create a realistic attack on the SEV until we reveal the physical address-based tweak algorithm used by SME and SEV. A linear function in Finite Field $F_2$ is used for the tweak algorithm. Given any two equal cipher-texts in two different physical addresses, we could conduct the relationship of the two corresponding plaintexts. And given any two physical addresses, we could calculate the required relationship of plaintext to generate equal cipher-texts in the two addresses. This opens a door for the Chosen Plaintext Attack.

## 4. Physical Address-based Tweak

To analyze the physical address-based tweak algorithm, we used a Ryzen machine with SME support. The AMD SME patch[20] was applied and SME feature was enabled. The SME kernel patch provided a function to switch the C-bit of a memory page. The function flushes cache before changing the C-bit and flushes TLB again after changing the C-bit. A new kernel module was created to provide the interface to change the C-bit for user level code.

When we write data into a memory page with C-bit unset and read data back from the memory page with C-bit set, the written data will be treated as cipher-texts and be decrypted automatically when it's read with C-bit set. On the other hand, if we write data into the memory page with C-bit set and read data back from the memory page with C-bit unset, the written data will be treated as plaintext and be encrypted automatically when it is flushed from cache to memory. After we read it again with C-bit unset, we could get the corresponding cipher-texts.

In our first experiment, we wrote plaintexts of arithmetic sequence into same physical address and read the corresponding cipher-texts sequence. A NIST randomness test suite[21] was used to verify that the output cipher-texts sequence had perfect randomness so that the mode of operation is not CTR but ECB..

Next we wrote equal plaintexts into different physical addresses in the memory page and found that the output cipher-texts also looked like a perfect random sequence.

| variable | value in hexadecimal |
|---|---|
| $t_4$ | 82 25 38 38 82 25 38 38 82 25 38 38 82 25 38 38 |
| $t_5$ | ec 09 07 9c ec 09 07 9c ec 09 07 9c ec 09 07 9c |
| $t_6$ | 40 00 00 18 40 00 00 18 40 00 00 18 40 00 00 18 |
| $t_7$ | 81 02 a2 3a 81 02 a2 3a 81 02 a2 3a 81 02 a2 3a |
| $t_8$ | 77 d9 10 77 77 d9 10 77 77 d9 10 77 77 d9 10 77 |
| $t_9$ | b0 10 b2 c0 b0 10 b2 c0 b0 10 b2 c0 b0 10 b2 c0 |
| $t_{10}$ | 53 6d 54 4d 53 6d 54 4d 53 6d 54 4d 53 6d 54 4d |
| $t_{11}$ | 15 68 ee 53 15 68 ee 53 15 68 ee 53 15 68 ee 53 |
| $t_{12}$ | b0 92 30 c2 b0 92 30 c2 b0 92 30 c2 b0 92 30 c2 |
| $t_{13}$ | 96 70 ff 8e 96 70 ff 8e 96 70 ff 8e 96 70 ff 8e |
| $t_{14}$ | 36 1b 90 d5 36 1b 90 d5 36 1b 90 d5 36 1b 90 d5 |
| $t_{15}$ | 04 00 c2 36 04 00 c2 36 04 00 c2 36 04 00 c2 36 |
| $t_{16}$ | e8 18 29 85 e8 18 29 85 e8 18 29 85 e8 18 29 85 |
| $t_{17}$ | bd 31 f9 2a bd 31 f9 2a bd 31 f9 2a bd 31 f9 2a |
| $t_{18}$ | a5 0d 37 44 a5 0d 37 44 a5 0d 37 44 a5 0d 37 44 |
| $t_{19}$ | f4 31 d8 4c f4 31 d8 4c f4 31 d8 4c f4 31 d8 4c |
| $t_{20}$ | 02 04 31 81 02 04 31 81 02 04 31 81 02 04 31 81 |
| $t_{21}$ | b3 71 32 a1 b3 71 32 a1 b3 71 32 a1 b3 71 32 a1 |
| $t_{22}$ | 50 8a c0 6c 50 8a c0 6c 50 8a c0 6c 50 8a c0 6c |
| $t_{23}$ | 16 8a 80 20 16 8a 80 20 16 8a 80 20 16 8a 80 20 |
| $t_{24}$ | 7f 9b c0 07 7f 9b c0 07 7f 9b c0 07 7f 9b c0 07 |
| $t_{25}$ | 00 db 04 07 00 db 04 07 00 db 04 07 00 db 04 07 |
| $t_{26}$ | 7f 00 04 04 7f 00 04 04 7f 00 04 04 7f 00 04 04 |
| $t_{27}$ | 70 fa 01 be 70 fa 01 be 70 fa 01 be 70 fa 01 be |
| $t_{28}$ | bb 3d 28 90 bb 3d 28 90 bb 3d 28 90 bb 3d 28 90 |
| $t_{29}$ | bd 2d d5 26 bd 2d d5 26 bd 2d d5 26 bd 2d d5 26 |
| $t_{30}$ | 1c 5d 6c e2 1c 5d 6c e2 1c 5d 6c e2 1c 5d 6c e2 |
| $t_{31}$ | af 4c 8f a4 af 4c 8f a4 af 4c 8f a4 af 4c 8f a4 |
| $t_{32}$ | 4f 5c e7 27 4f 5c e7 27 4f 5c e7 27 4f 5c e7 27 |
| $t_{33}$ | af 4c 8f a4 af 4c 8f a4 af 4c 8f a4 af 4c 8f a4 |

Table 1. Tweak function parameters

Finally we tried to write equal cipher-texts into different physical addresses in the memory page and found the output plaintexts could not pass the randomness test. Manual analysis on the first several lines of output plaintexts showed that the data should be in linear relationship in finite field $F_2$ and the linear relationship could be verified by a C code.

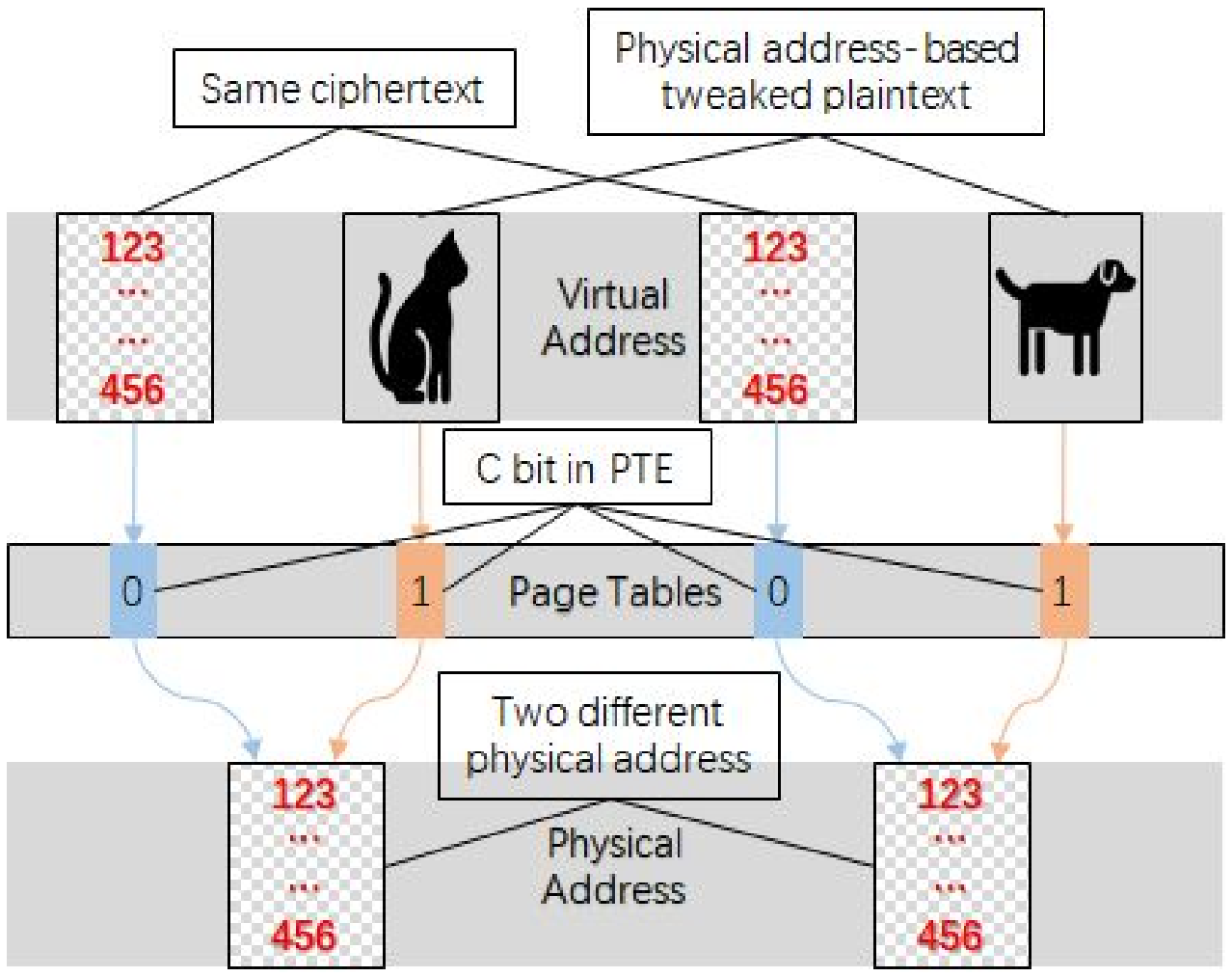

Figure 8. Equal ciphertexts in different addresses

We then randomly allocated many memory pages with C-bit unset and wrote equal cipher-texts into all of those memory pages. Setting C-bits of all those memory pages and reading back all those plaintexts, we got a list of plaintexts together with corresponding physical addresses with equal cipher-texts. Using Gaussian Elimination in the field $F_2$, we found out that the physical address-based tweak function is a linear function with parameter as Table 1.

We could define a function $T(x) = \oplus_{x_i=1} t_i$ , where $x_i$ is the $i$ th-bit of integer $x$, $\oplus$ is the sum of vectors in finite field $F_2$ (or bitwise exclusive-or) and $t_i$ is a 128-dimensions constant vector in $F_2$ as in the table 1. For plaintext $m_1$ in address $p_1$ and plaintext $m_2$ in address $p_2$ , the ciphertext is equal if and only if $m_1 \oplus T(p_1) = m_2 \oplus T(p_2)$ or $m_2 = m_1 \oplus T(p_1 \oplus p_2)$ since $T$ is linear function, where $\oplus$ is the bitwise exclusive-or operation.

## 5. Simulate Attack of SEV

Since there is no integrity protection provided by SEV and malicious hypervisor is able to modify the cipher-texts of VM, with tweak function $T(x)$ provided in section 4, malicious code in hypervisor is able to inject any code into any address of a VM. In this section, we show how the attack could be done efficiently.
Since SEV enabled machine is still not available in market, the attack is simulated in a Ryzen machine which supports SME. We believe the attack does also work in SEV enabled machine since the same encryption algorithm is used by both SME and SEV according to AMD's *Memory Encryption Whitepaper*.
It is assumed that hypervisor has been compromised so that the attack is started from privileged code of hypervisor.

## 5.1 Prerequisite

Both host and guest VM use Ubuntu 16.04 in our demo. We assume that the OS image of both host and guest VM is public available.
We also assumed that the guest VM needs to communicate with hypervisor or other external parts so that an external part could send some data into the VM and part of the data will be kept in the memory of VM, and encrypted automatically by guest VM VEK. For example, guest VM may receive data from external part by network or guest VM may read data in a virtual hard disk provided by hypervisor. By assuming that the attacker has controlled the hypervisor and host OS, the attacker has the ability to apply the Chosen Plaintext Attack by sending chosen plaintext to the VM.
A popular http server-- nginx is pre-installed in the guest VM to work as a data receiver in the VM so that hypervisor could deliver data to VM via nginx.
We also need a victim code with root privilege in the VM. An ssh server-- sshd is pre-installed and enabled in the guest VM as the victim code. Similar attack could be applied to kernel codes or other code with root privilege too.

## 5.2 Driver to simulate SEV environment

The basic idea is that we start a VM with all contents to be encrypted (so C-bits in corresponding page table are all set). At the same time, all the memory pages of VM is remapped by a new page table in hypervisor with all C-bits unset and the code to simulate the attack should only read encrypted-data of VM via the new page table.
QEMU[22] and KVM[23] are used in our demo to start a VM. SME is enabled in the demo machine so that when started, C-bits in all nPT entries are set so all memories used by the VM are encrypted automatically. To achieve the goal to simulate the encrypted VM and hypervisor could read cipher-texts only, we created a driver module in the host OS and we name it the assist driver. The assist driver provides:
i) Scan nPT of the VM (which is visible to hypervisor) and find all physical memories assigned to the VM
ii) Create new page table entries in the hypervisor to map all physical memories assigned to VM into virtual address of hypervisor and unset C-bits of all those pages.
iii) The hypervisor is able to read and write cipher-texts of all memory pages owned by VM (cache coherence instructions are required).

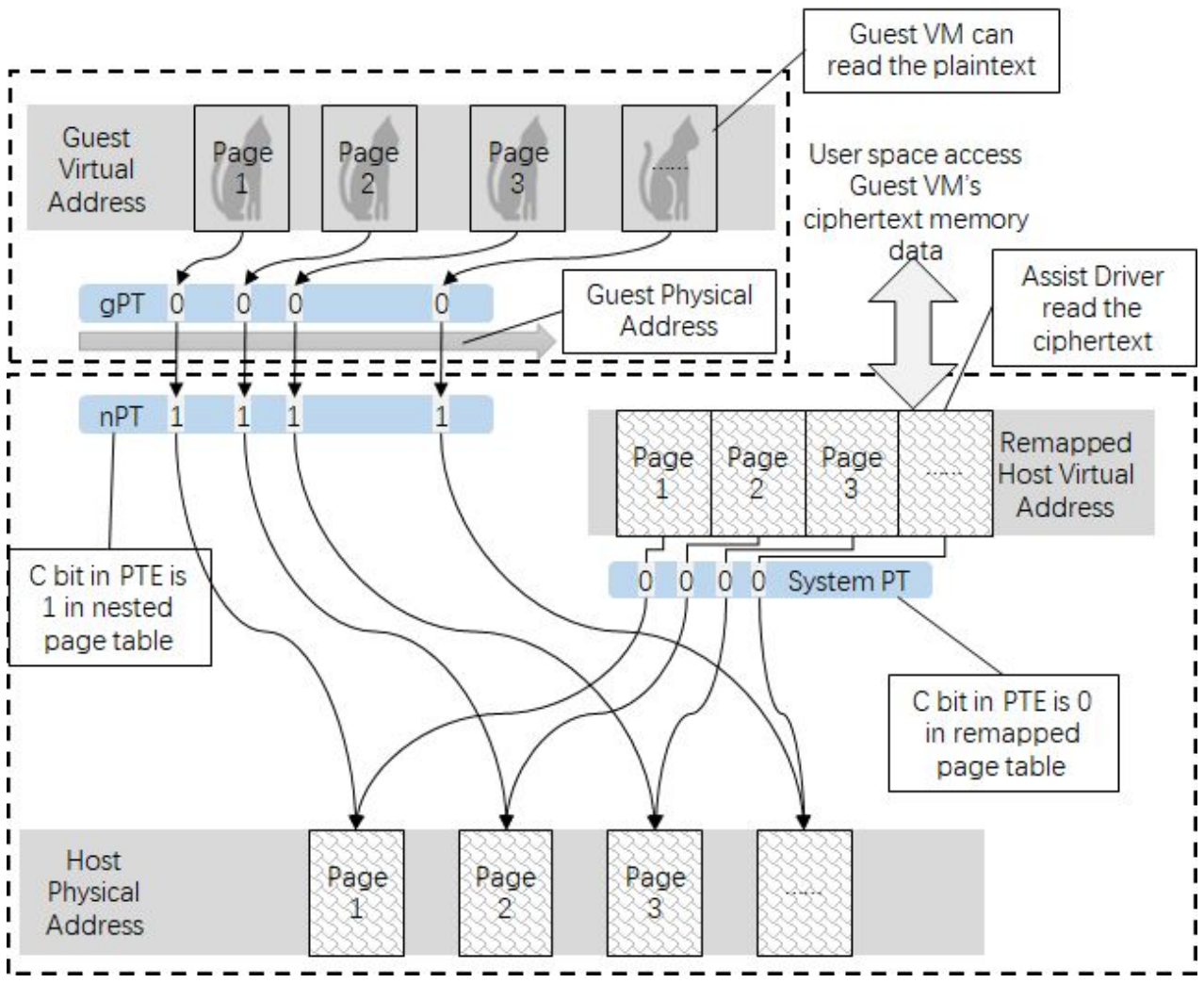

Figure 9. Remap address of VM with C-bit unset

## 5.3 Prepare data receiver in VM

A popular http server--nginx is pre-installed in the guest VM to play the role as data receiver inside the VM. First the http server is tested in a VM without memory encryption. We found that Mempool is used in nginx for memory management. When an external user sends some data to the server, part of the data is almost always left in a fixed memory page as long as the server is not restarted.

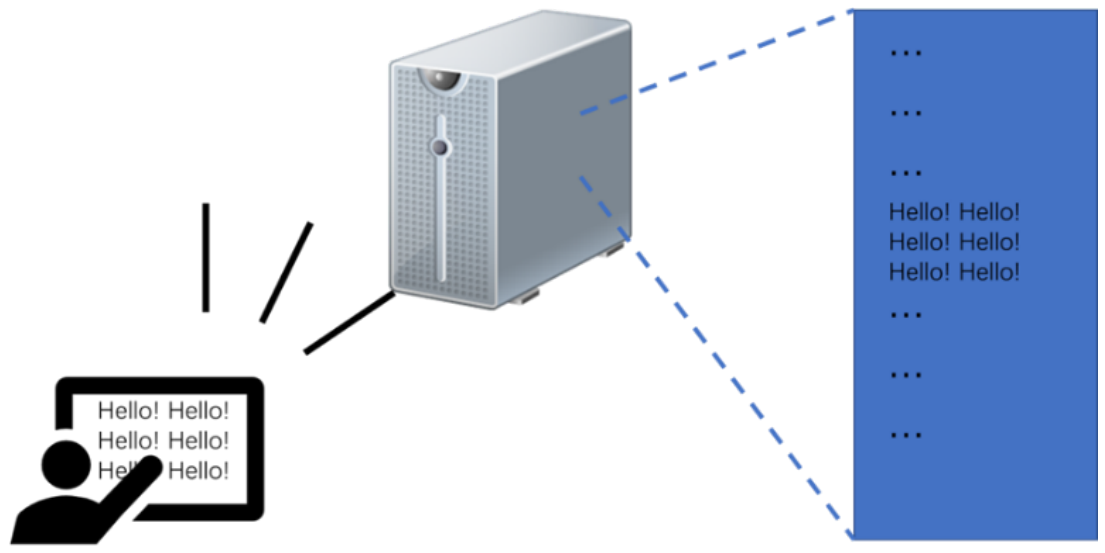

Figure 10. Send data to VM

We noticed although some of data sent to the server is overwritten after the server finish processing the data, at least $62 \times 16$ bytes of injected data is left in the memory page. We called the memory region to receive the $62 \times 16$ bytes of injected data as Bridge and it provided the opportunities for Chosen Plaintext Attack. We then called the memory page containing the Bridge as Bridge Page. We found that the offset of Bridge inside the Bridge Page never changes even if the server was restarted.

We could use any http client to inject data into the Bridge, such as command

*$nc 192.168.1.2 80 < target_file.txt >*

injected data in file target_file.txt into the Bridge. We have also tried other http clients, such as browser with GUI, curl or tool using TCP/Socket, all of them could successfully inject data into Bridge.

## 5.4 Prepare the Victim

We selected sshd as the victim code because it's a popular tool and runs in root privilege and it provides user connected to the service a shell interface so that it is convenient for demo.

First we prepared a 48 bytes code (which is named as shellcode) to fork a child process running as root and provide a shell interface to attack. We also found the code in sshd to authenticate the user login information and the goal of the attack is to replace the authentication code by the 48 bytes shellcode. We called the data of the first 16 bytes of the code to be replaced as the Characteristic Code.

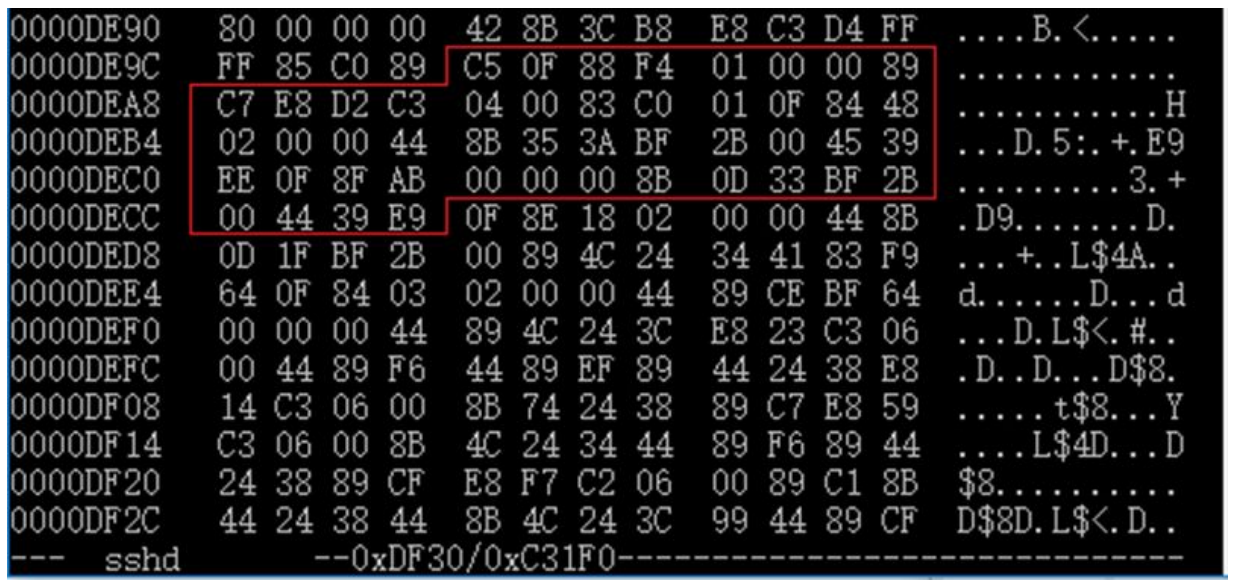

Figure 11. The 48 bytes Victim code

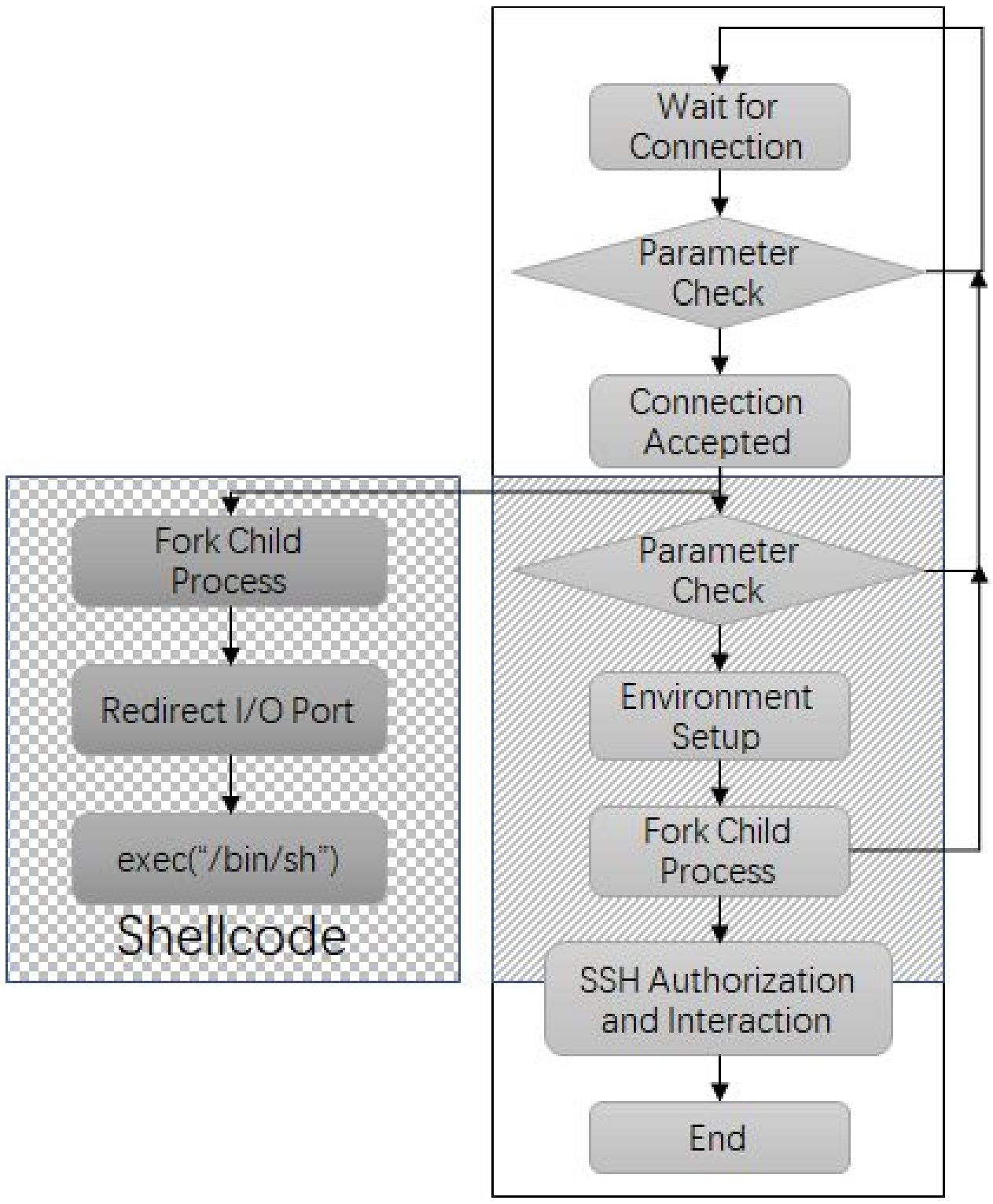

Figure 12. Logic of victim and injected code

## 5.5 Search for Bridge

Since the address of Bridge changes every time the http server is restarted, we needed to find the physical address of a Bridge in a VM in hypervisor.

The Bridge contains 62 groups of continuous 16 bytes data. Let's assume the offset of the 62 groups of data in the Bridge Page to be $o_1, o_2 = o_1 + 16, \ldots, o_{62} = o_1 + 976$. If we send data $T(o_1), T(o_2), \ldots, T(o_{62})$ into the Bridge where function $T(x)$ is defined in section 4, the cipher-texts of all the 62 groups should be exact same according to result of section 4 taking advantage of the factor that function $T(x)$ is a linear function (If $T(x)$ is enhanced to a nonlinear function, even if physical address is tweaked into both plaintext and cipher-text, we could still find out the bridge by algorithm similar to section 5.6 although it requires longer time to search for the bridge).

So we could find the address of Bridge by:

i) Generate data $T(o_1), T(o_2), \ldots, T(o_{62})$.

ii) Send the above data by any http client to the Bridge which will be encrypted by hardware using guest VM VEK automatically.

iii) Dump out cipher-texts in all physical memory pages of VM by assist driver.

iv) Analyze above ciphertext to find 62 groups continuous equal ciphertexts to identify physical address of Bridge -- the BPA.

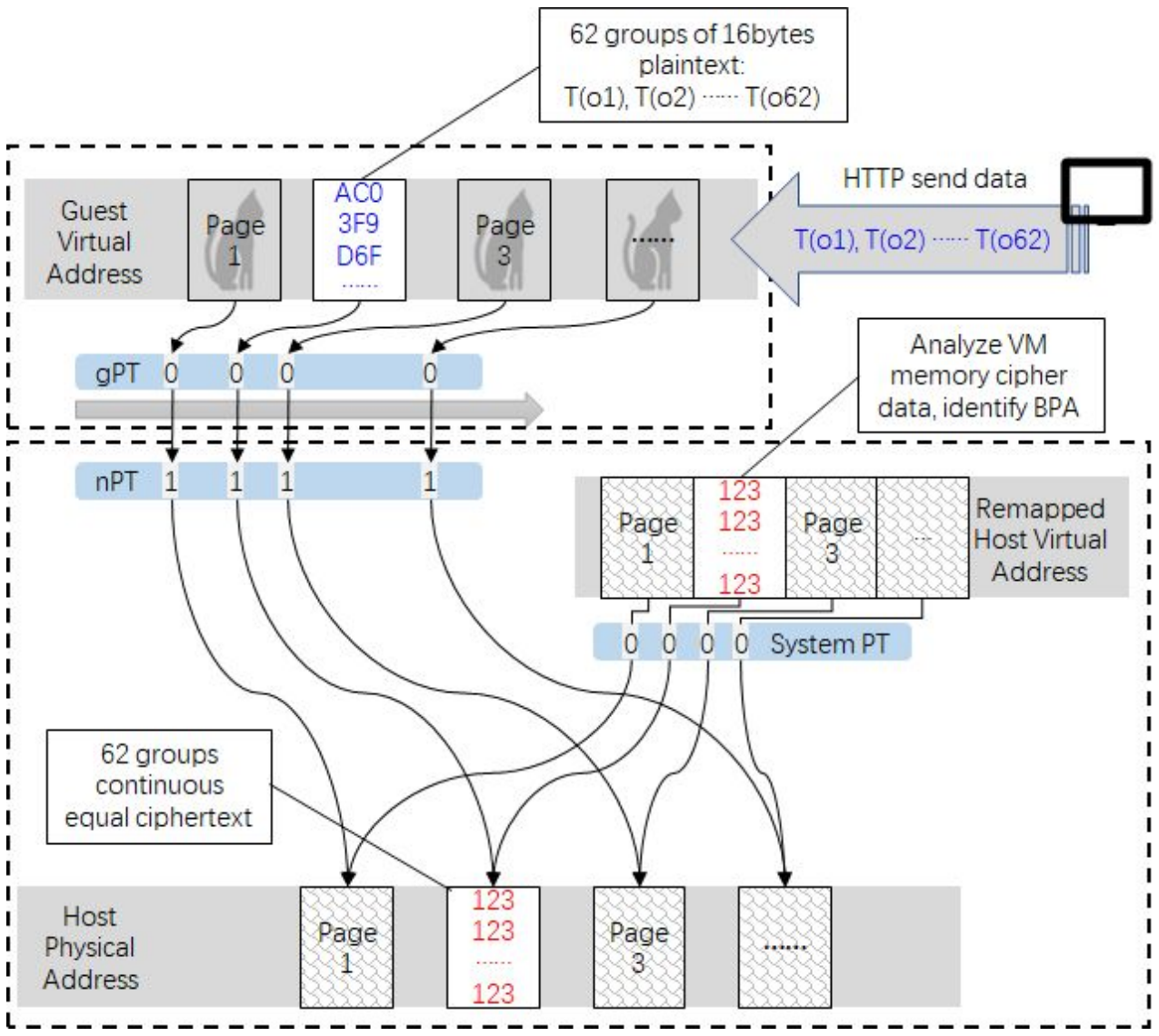

Figure 13. Logic to Search for Bridge

## 5.6 Search for Characteristic Code of Victim

We need also to determine the physical address of the Characteristic Code in the code of victim sshd. The below steps could be used to enumerate the physical address of Characteristic Code:

1. We have found physical address of Bridge BPA in section 5.5

2. We also have the 16 bytes Characteristic Code: $CC$ and the offset inside page is known

3. For every 16 bytes physical address $CCPA_CAND$ whose offset inside page is same as that of CC

3.1. Calculate $CCG = CC \oplus T(BPA \oplus CCPA_CAND)$

3.2. Send 16 bytes $CCG$ to Bridge by http client

3.3. Read and compare cipher-texts in $BPA$ and $CCPA_CAND$ by assist driver. The Characteristic Code is in physical address $CCPA_CAND$ if the two cipher-texts match.

4. End For

Reading of cipher-texts by assist driver requires deactivating of guest VM and flushing cache and TLB and it takes a lot of time. Since there're total 62 groups of 16-bytes of data in the Bridge, we could improve the above algorithm searching for 62 physical address candidates of CC in each round.

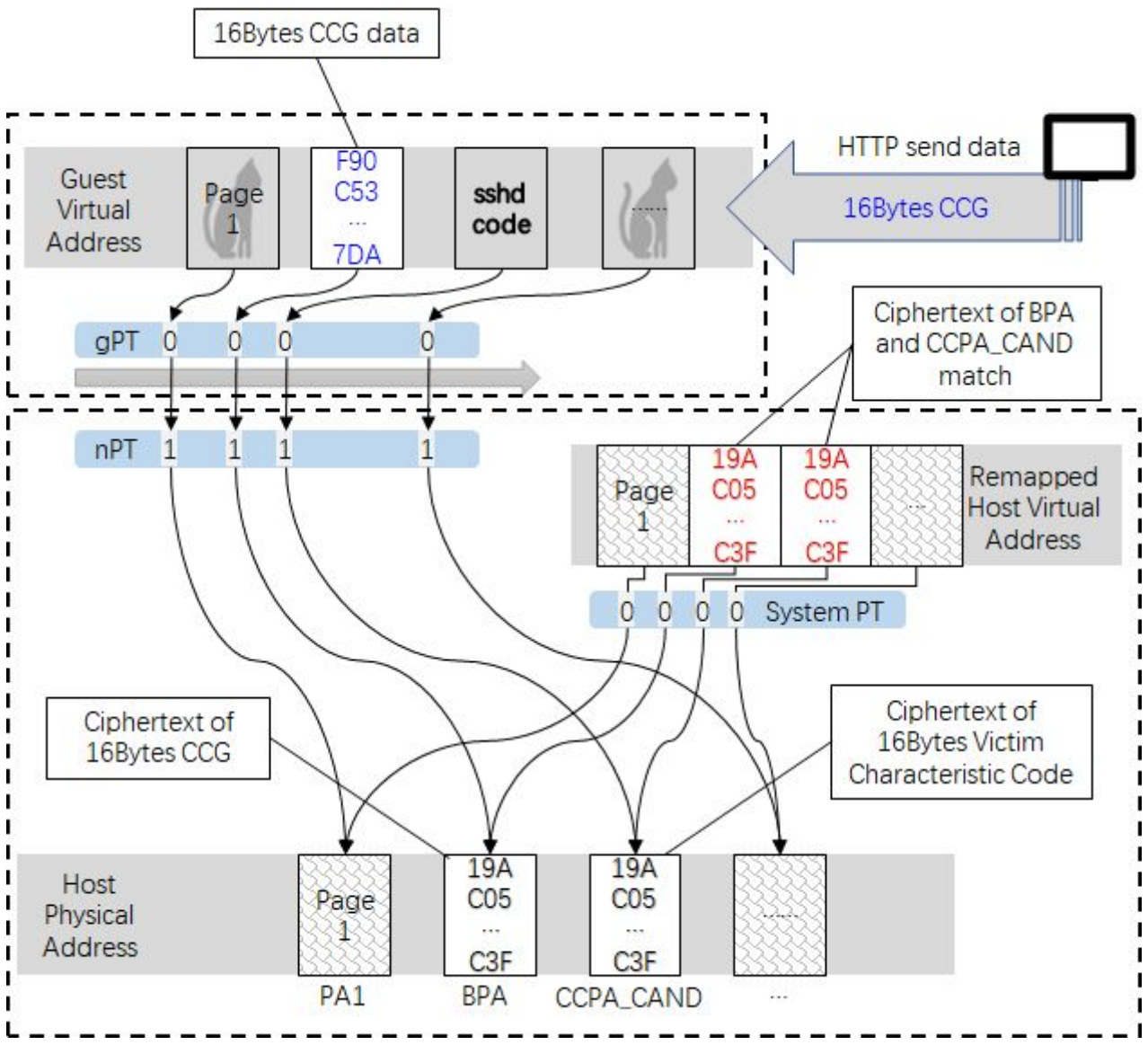

Figure 14. Logic to Search for Characteristic Code

## 5.7 Inject Shellcode

Now both BPA and physical address of CC (PCC) have been available. The next step is to change the code in PCC to Shellcode. What we need do is:

i) Calculate 48 bytes data

$$T(PCC \oplus BPA) \oplus Shellcode[1\sim16]$$

$$T((PCC + 16) \oplus (BPA + 16)) \oplus Shellcode[17\sim32]$$

$$T((PCC + 32) \oplus (BPA + 32)) \oplus Shellcode[33\sim48]$$

ii) Send above data to Bridge by http client

iii) Read the first 48 bytes of cipher-texts in Bridge by assist driver

iv) overwritten cipher-texts of the first 48 bytes in PCC by the above cipher-texts

After that, the shellcode has been injected into the victim sshd.

Next, we could connect to the ssh server from hypervisor to obtain the root privilege of the guest VM.

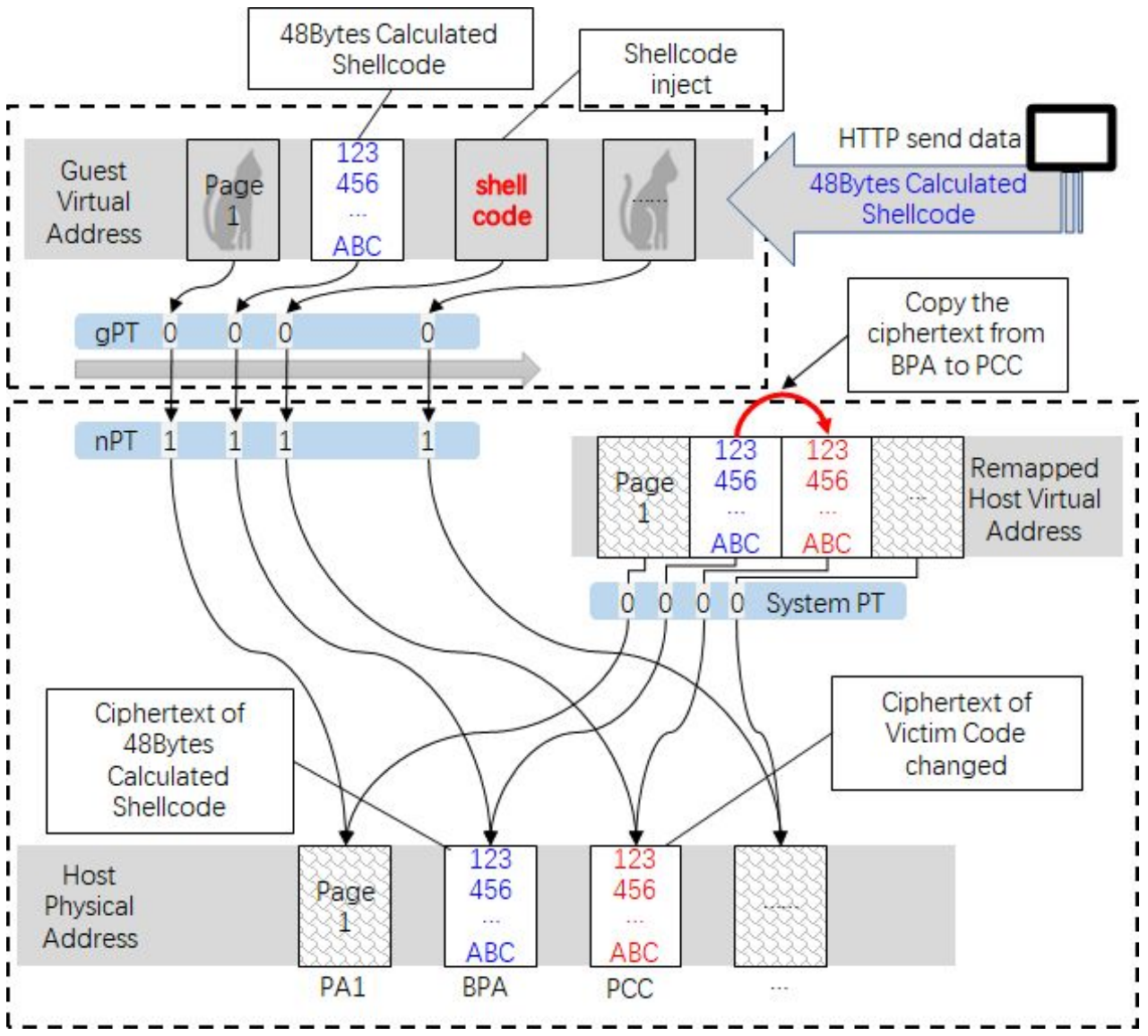

Figure 15. Overwritten Characteristic Code by ShellCode

## 6 Result

The demo is done with an HP machine whose CPU model is AMD Ryzen 7 1700X with both SVM and SME enabled. Ubuntu 16.04.3 LTS is installed as host OS and the system memory is 8GB DDR4 2400MHz. A VM is started from Qemu and the memory size of the VM is 1GB. The OS Ubuntu 16.04.3 LTS is also used by VM and OpenSSH_7.2p2 Ubuntu-4ubuntu2.2 and nginx/1.10.3 are installed in the VM.

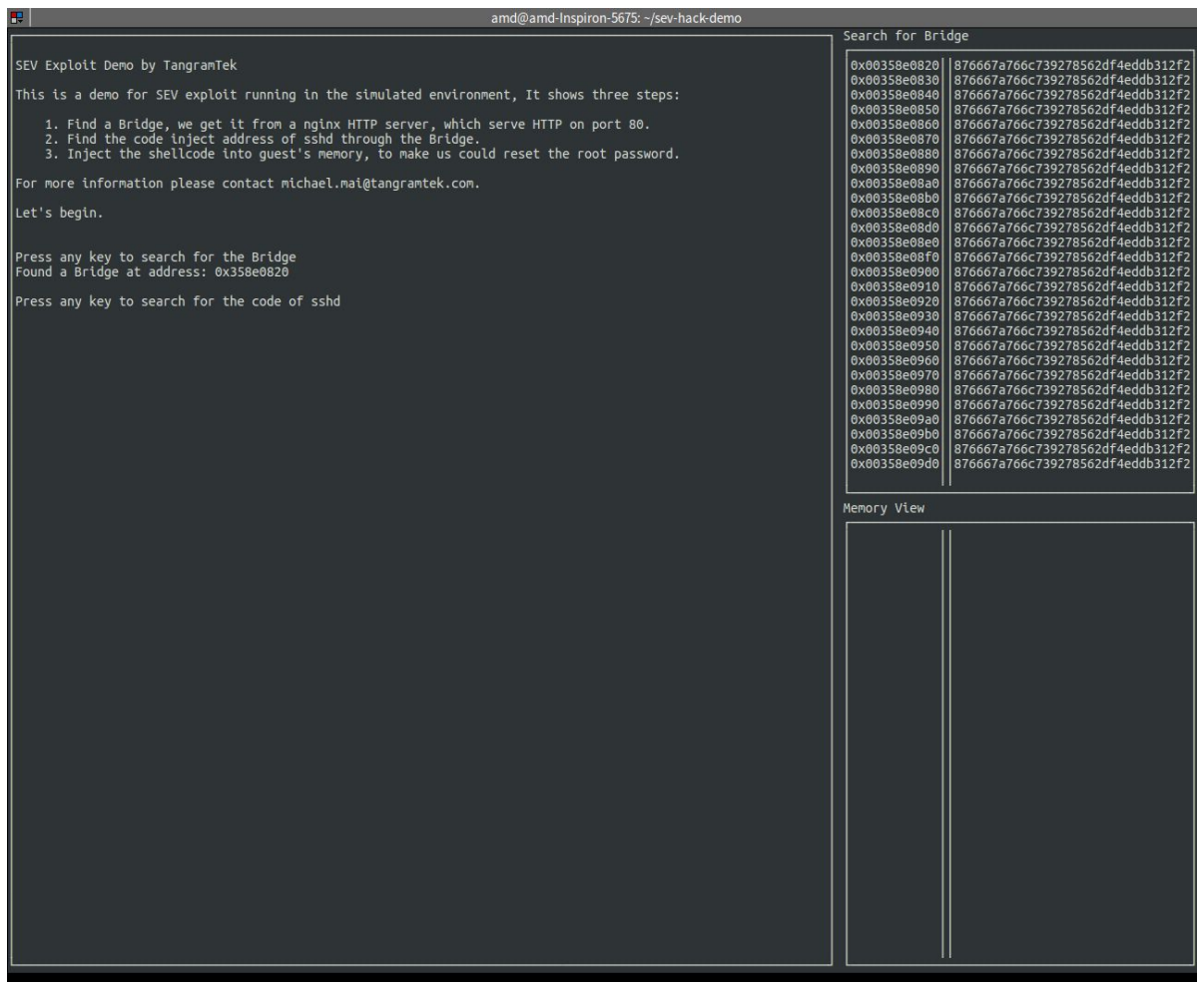

Figure 16. Search for Bridge

The time to search for Bridge and inject shellcode to replace Characteristic Code could be ignored and most time is consumed in searching for Characteristic Code. It takes about 20 seconds on average to find the Characteristic Code.

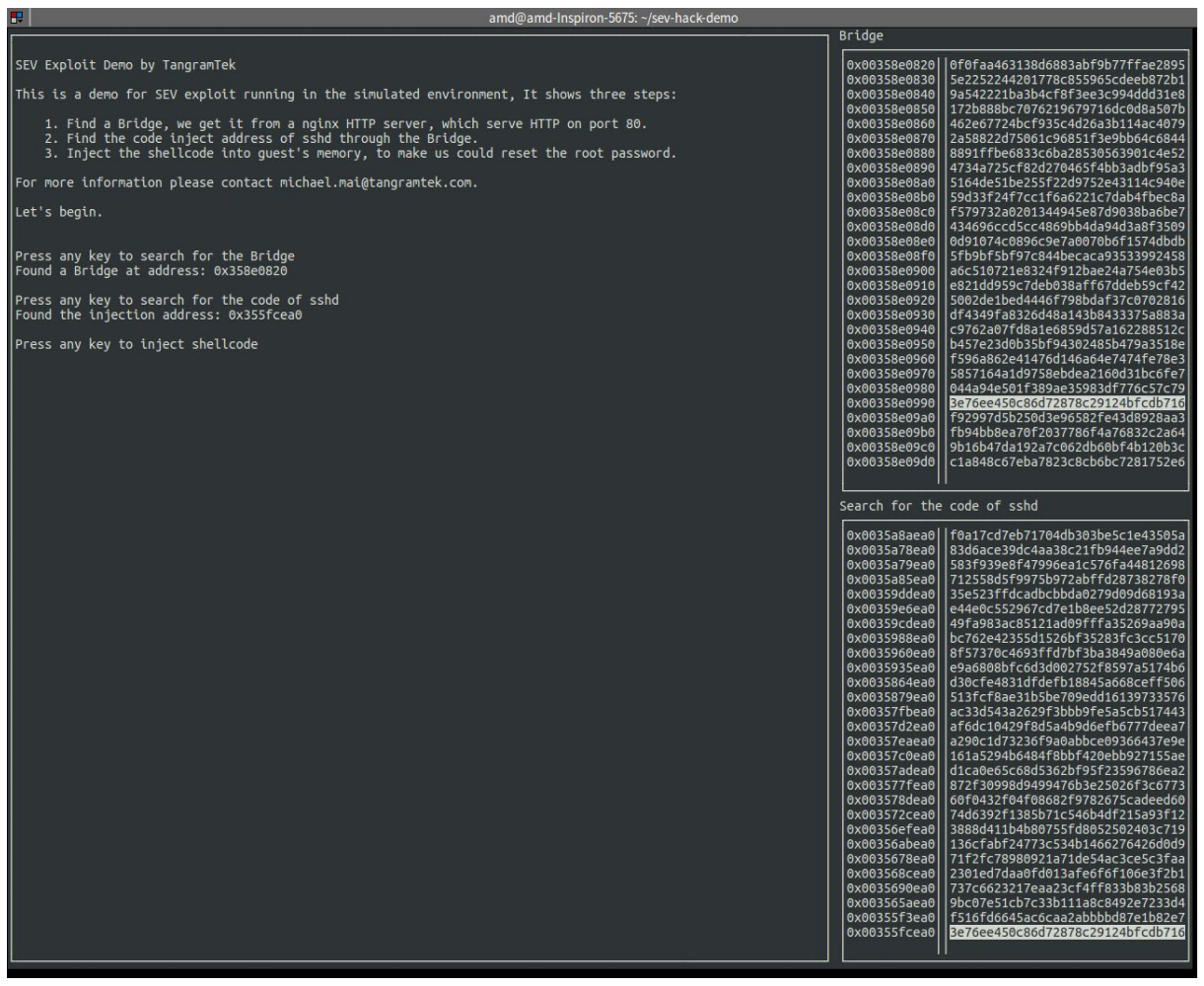

Figure 17. Search for Characteristic Code

## 7. Conclusions

In this paper, we have demonstrated that the current implementation of SEV is vulnerable. We suggest AMD to update the physical address-based tweak algorithm before releasing SEV into market. The physical address-based tweak algorithm should not tweak the address into plaintexts or cipher-texts. The address should be tweaked into key of AES algorithm to protect against cipher-text move attacks. It is preferred that Key Derivation Function such as the one specified in NIST SP 800-108[24] should be used to tweak the address into the key. Encryption is not enough to isolate VMs from hypervisor and it is better that integrity protection could also be provided. The idea to provide a technology to isolate data of guest VM from hypervisor is promising, but there're still a lot of improvement opportunities in the current implementation.